\documentclass[aps,prl,preprint,showpacs]{revtex4}
\usepackage{graphicx}

\begin{document}

\title{Non-equilibrium coherence dynamics of a soft boson lattice}
\author{A. K. Tuchman$^1$, C. Orzel$^2$, A. Polkovnikov$^3$,
M. A. Kasevich$^1$}
\affiliation{$^1$Physics Department, Stanford University, Stanford CA, 94305}
\affiliation{$^2$Physics and Astronomy Department, Union College, Schenectady, NY 12308}
\affiliation{$^3$Physics Department, Harvard University, Cambridge, MA 02138}
\date{\today}

\begin{abstract}
We study the non-equilibrium evolution of the phase coherence of a
Bose-Einstein condensate (BEC) in a one dimensional optical
lattice, as the lattice is suddenly quenched from an insulating to
a superfluid state.  We observe slowly damped phase coherence
oscillations in the regime of large filling factor ($\sim$100
bosons per site) at a frequency proportional to the generalized
Josephson frequency.  The truncated Wigner approximation (TWA)
predicts the frequency of the observed oscillations.
\end{abstract}
\pacs{03.75.Kk,03.75.Lm}
 \maketitle

Proposals for interferometry with BEC have demonstrated that
number squeezed states can potentially provide robust
sub-shot-noise sensitivity to perturbing interactions
\cite{Burnett, Mark, Burnett2}. The appeal of one recent proposal
by Dunningham and Burnett (DB) \cite{Burnett3} is
the current availability of number squeezed states in an optical lattice \cite%
{Orzel,Bloch} and the simplicity of the required experimental
sequence.  In this proposal, an array of number squeezed states is
initially prepared in a suitably deep optical lattice.  The
lattice depth is then rapidly lowered and the system is allowed to
evolve in the presence of a perturbing potential energy gradient.
The lattice intensity is subsequently restored to its initial
value. A final interferometric measurement of array phase
coherence is used to characterize the perturbation. This sequence
is loosely analogous to sub-shot-noise optical interferometry
using Fock states as inputs to a Mach-Zehnder interferometer
\cite{Burnett}.

In this work, we examine the first stage of the DB interferometer
by studying the diabatic response of highly squeezed number states
in an optical lattice to a sudden reduction in the lattice
intensity.  This sequence induces on-site phase variance
oscillations as the quantum state evolves.  We observe these
oscillations through characterization of the dynamic evolution of
interference contrast.

Unlike the optical case, analysis of the lattice interferometric
sequence is complicated by the presence of a strong non-linearity
and multiple interfering modes. For the parameters of our
experiments (many atoms and many lattice sites), exact solutions
of the many-body equations of motion are unavailable due to the
exponentially large Hilbert space needed to describe the system.
Furthermore,  traditional approximations fail due to their
inability to describe the initial state \cite{Tolya}.  Thus,
analysis of this dynamic evolution is an interesting problem in
its own right.

The Bose-Hubbard Hamiltonian accurately describes the
atom--lattice system \cite{Fisher,Sachdev,Jaksch}.  Written in
terms of single particle creation and annihilation operators
$\hat{a}_{i}^{\dagger }$ and $\hat{a}_{i}$,

\[
H = -\gamma{\sum_{<i,j>}}\hat a_{i}^{\dagger}\hat a_{j} + {}\frac{1}{2}%
g\sum_{i} \beta_i\hat a_{i}^{\dagger}\hat a_{i}^{\dagger}\hat
a_{i}\hat a_{i} +\sum_{i}{V_i}\hat a_{i}^{\dagger}\hat a_{i},
\hspace{1cm}(1)
\]
where the subscripts index the lowest vibrational mode for each
lattice site.  Here $\gamma$ is the tunneling rate between
adjacent lattice sites, $g\beta_i$ is the mean field energy due to
repulsive interactions between two atoms ($g=4 \pi \hbar^2 a/m$,
$a$ is the s-wave scattering length, and $m$ is the atomic mass),
$N_i$ the number of atoms per site, and $V_i = \Omega i^2$ is the
external potential due to a harmonic magnetic trap. $\beta_i$
(which depends on the on-site boson occupancy) and $\gamma$ are
determined from integrals over single particle wavefunctions
\cite{integrals}. The importance of quantum fluctuations is
determined by the ratio $g\beta_i/N_i \gamma$, where
$g\beta_i/N_i\gamma \sim 1$ indicates the superfluid to
Mott-Insulator (MI) crossover \cite{Fisher,Jaksch,Zwerger}. We
characterize global array phase coherence by the quantity $D(t) =
\sum_{i\neq j}\langle a_{j}^{\dagger }a_{i}\rangle/M $, where $M$
is defined as the ratio of the total number of atoms to the
number of atoms in the central lattice site \cite{Tolya}.

\begin{figure}[htb!]
\includegraphics[scale = .5, bb = 100 60 500
630]{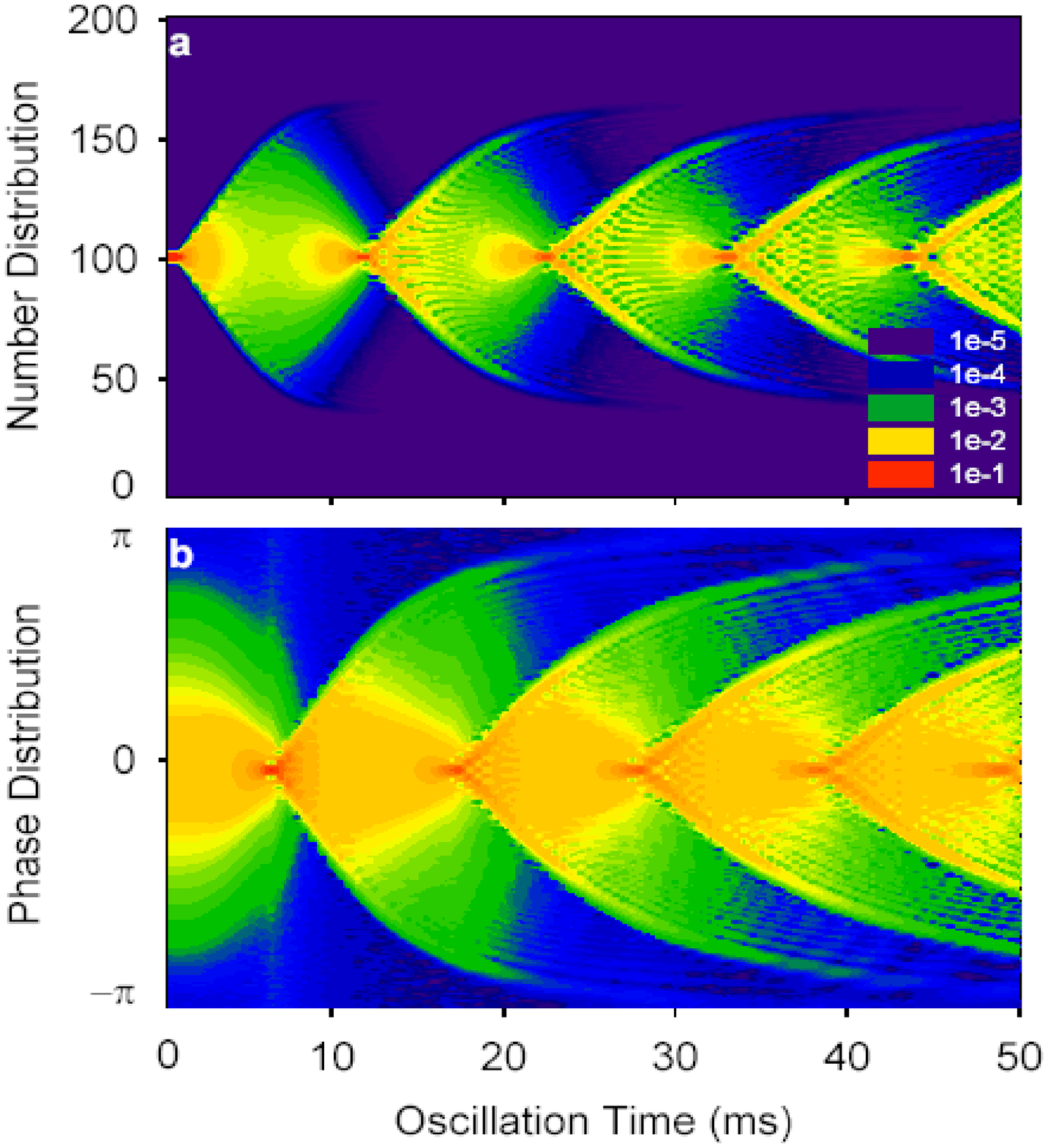} \caption{ \newline {\protect\small a) Evolution of
the number distribution of a two site system following a 200 ms
lattice intensity ramp to $65$ $E_R$ and a subsequent diabatic
lowering of the tunnel barrier to $20$ $E_R$, for N = 100 atoms
per lattice site. False color exponential scaling denotes
probability of number distribution in one site. b) Associated
phase distribution. }} \label{pofig1}
\end{figure}

Qualitative understanding of the system dynamics can be obtained
by solving Eq. 1 for a two lattice site model. Fig.~\ref{pofig1}a
illustrates the evolution of the lattice number distribution
following a sudden reduction in lattice intensity under conditions
similar to those used in the experiments described below.
Fig.~\ref{pofig1}b shows the associated phase distribution
(obtained through a Fourier transform of the number distribution
coefficients).  For the two site system, the characteristic
frequency for these oscillations is determined by the generalized
Josephson frequency $\sqrt{4N_i g \beta_i \gamma+4\gamma^2}$
\cite{smerzi2,Legget}.  At the point of the first phase revival
($t$ $\sim$ 6 ms in Fig. 1b), the phase variance is sub-Poissonian
while the number variance is super-Poissonian.  In principle, this
enables interferometric measurement of phase shifts below the atom
shot-noise limit \cite{Mark}.

In the case of the full lattice system, we employ the TWA ~\cite
{Steel,ap1} to obtain approximate array dynamics.  Under this
approximation, the quantum mechanical expectation value of an
observable is replaced by a semiclassical ensemble average.
Specializing to the lattice system, we consider a set of
wavefunctions $\psi_j$ which evolve according to the
semiclassical, discrete Gross-Pitaevskii equation of motion (GPE)
\cite{GPeq,Tolya}:

\[
i\hbar \frac{\partial \psi _{j}}{\partial t}=-\gamma (\psi _{j-1}+\psi
_{j+1})+(V_{j}+g{\beta_j}{\mid }{\psi _{j}}\mid ^{2})\psi _{j}.\hspace{1cm%
}(2)
\]%
The initial condition $\psi _{j}(0)=\sqrt{N_{j}}e^{i\phi _{j}}$,
where the phases $\phi _{j}$ are sampled from a uniformly
distributed ensemble of values between $0$ and $2\pi $, and $N_j$
are determined from the GPE groundstate solution.

The TWA approach is accurate for suitably short times and can be
applied to this problem since the time evolution occurs in the
semiclassical superfluid regime \cite{ap1}.  It has been used to
analyze the breakdown of adiabaticity for the lattice squeezing
experiments described in Ref. \cite{Orzel} and also to study
damping of dipolar motion in an optical lattice \cite{ruos,pw}.
Density matrix renormalization group (DMRG) techniques have
recently been used to study non-equilibrium dynamics of boson
lattices with low filling factor \cite{Clark}, but these methods
are difficult to extend to large filling factors \cite{Vidal}.

The apparatus used in this work is described in
Refs.~\cite{Anderson, Orzel}.   $^{87}$Rb atoms confined in a TOP
trap are evaporatively cooled to produce nearly pure condensates
in the $F=2, m_F=2$ state which contain $ \sim 3 \times 10^{3}$
atoms at $T/T_c < 0.3$.  After adiabatically relaxing the
confining harmonic potential, the BEC is loaded into a vertically
oriented one dimensional optical lattice.  The lattice is produced
by a retro-reflected $\lambda = 840\,$nm laser beam focused to a
$50\,\mu$m $1/e$ intensity radius at the point of overlap with the
BEC.  This light field provides strong transverse confinement in
addition to periodic confinement along the propagation axis. For
example, at a well depth of $U =63 \, E_R$ ($E_R/\hbar = \hbar
k^2/2m \sim 2\pi \times$ 3.23 kHz, with ${k} = 2\pi/\lambda$), the
transverse oscillation frequency is $150$ Hz, significantly larger
than the $11$ Hz radial frequency of the magnetic trap
\cite{lattice}.

We load the atoms into the optical lattice by linearly increasing
the lattice intensity to $U_0$ in $200$ ms (see Fig. 2a).  At $U_0
= 63 \, E_R$, we infer $\gamma/\hbar = 2\pi$ $\times$ 0.019 Hz,
$N_0 = 90$ and $g \beta_0/N_0 \gamma = 2$ for the central lattice
site ($i = 0$).  We then rapidly lower the lattice intensity (in a
$2$ ms ramp) to a final level $U_f$ with a corresponding
groundstate in the superfluid regime. In our experiments, $U_f$
ranges from 12 to 32 $E_R$, covering the parameter intervals  1 Hz
$\le g\beta_0 / 2 \pi \hbar \le$  2 Hz and  42 Hz $\ge$ $\gamma/2
\pi \hbar$ $\ge$ 1 Hz. The time scale for lowering the potential
is chosen to be fast compared to the characteristic time scale for
adiabatic evolution of the many-body ground state but slow with
respect to the oscillation frequency of the individual wells.

\begin{figure}[htb!]
\includegraphics[scale = .50, bb = 50 50 600 740]{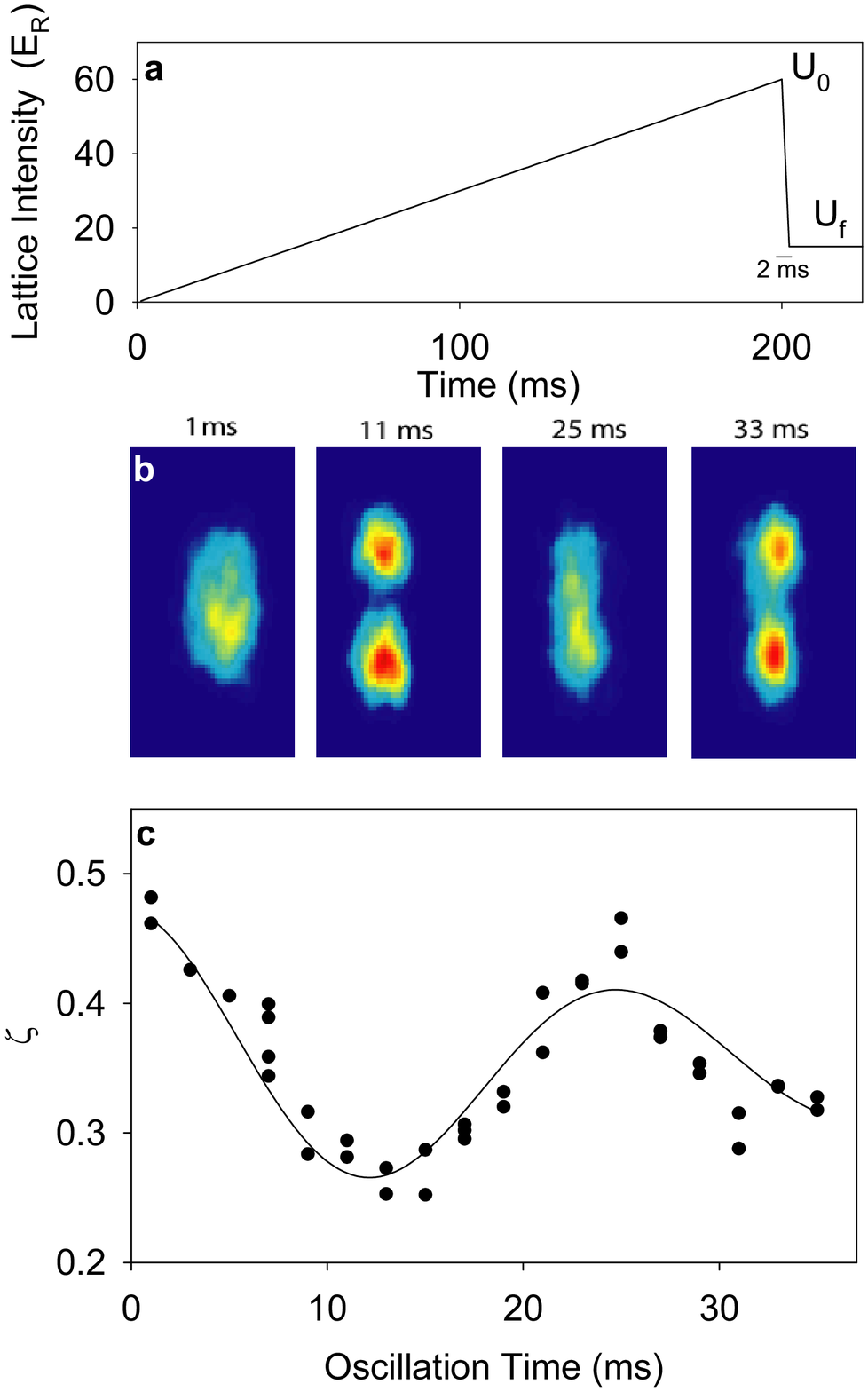}
\caption[Experimental Phase Variance Oscillations]{ \newline
{\protect\small a)Experimental sequence for the lattice intensity
ramp. The lattice is ramped to a peak value $U_0$ near the
insulating regime and then rapidly decreased to $U_f$. b)
Absorption images of the atom density profile for indicated hold
times. c)Phase
variance oscillation with $U_0 = 63 \,E_R$, $U_f = 16.6 \, E_R$ and $%
\Omega/\hbar = 2\protect\pi\times 0.67 \,$ Hz. }} \label{figure2}

\end{figure}

We stroboscopically follow the evolution of the array phase
coherence after lowering the lattice intensity to $U_f$ by holding
the atoms in the lattice at $U_f$ before releasing them and
observing the interference of their de Broglie waves.
Fig.~\ref{figure2}b displays a series of absorption images of the
atomic density profile as the quantum state evolves.  We define a
contrast parameter $\zeta$ as the ratio of the width of a single
peak to the separation between the peaks; large $\zeta$ indicates
loss of interference contrast.  Fig.~\ref{figure2}c shows the
oscillatory response of $\zeta$ as a function of hold time.
Remarkably, the system evolves from an initial state with poor
coherence to one with sharp coherence and then returns back,
possibly indicating a transition from an insulator to a
superfluid, then back to an insulator.

\begin{figure}[htb!]
\includegraphics[scale = .4, bb = 0 50 650 770]{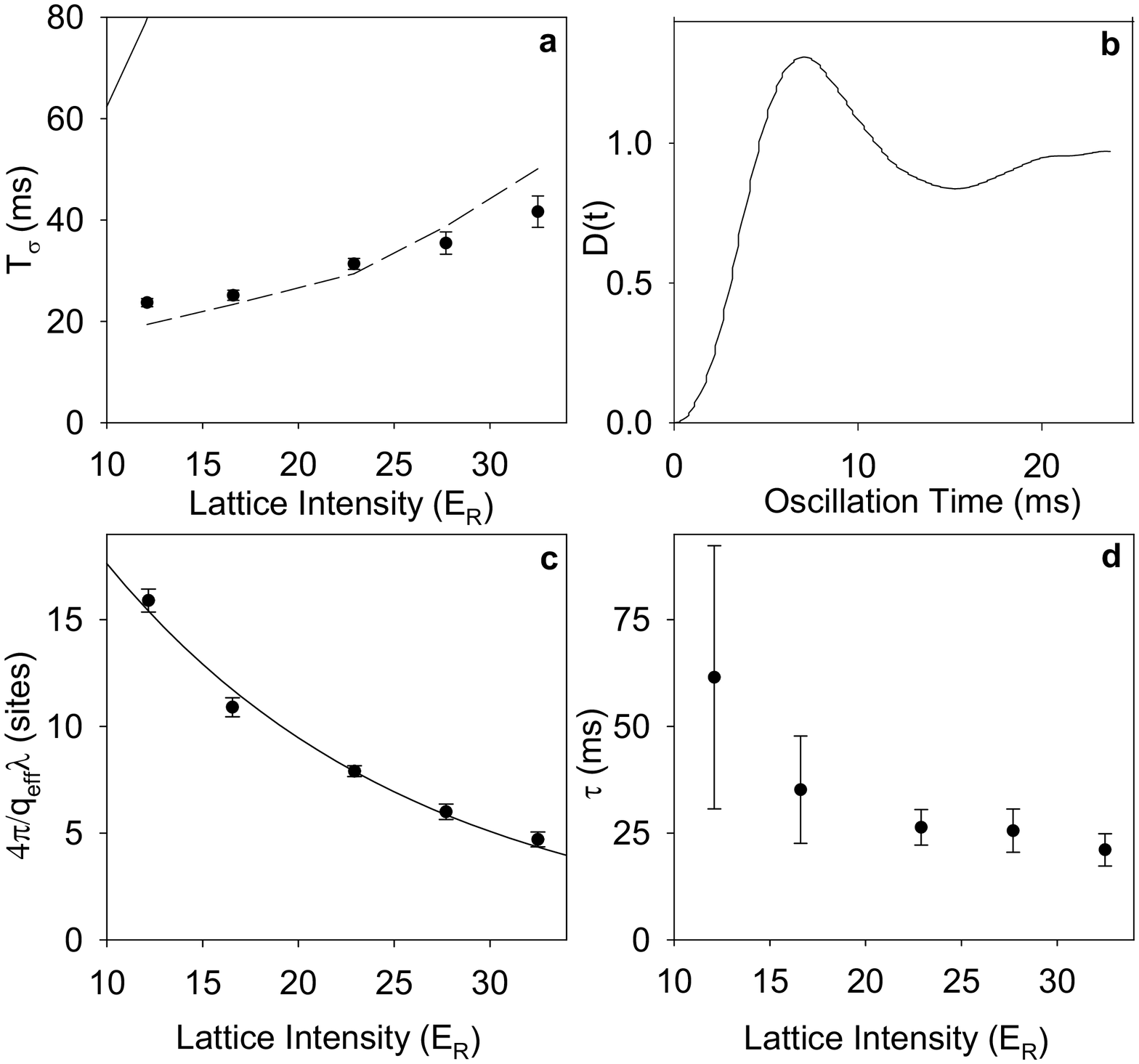}
\caption[Comparison with Theory] {\protect\\ \protect\small a)
Phase contrast oscillation period vs. lattice intensity. Dashed
line depicts results of TWA scaled by a factor of 2.1.  For
comparison, solid line in the upper left hand corner denotes the
quadrupole mode excitation spectrum calculated using GPE. b) TWA
simulation for D(t) with $U_f$ = 23 $E_R$. c) $4
\pi/q_\mathrm{eff}\lambda$ lattice sites vs. lattice intensity.
Solid line is exponential fit to data. d) Damping coefficient
$\tau$ vs. lattice intensity. }\label{figure3}
\end{figure}

To make contact with theory, we study the dependence of the phase
variance oscillation period $T_\sigma$ on the final lattice depth
$U_f$ (see Fig.~\ref{figure3}a). $T_\sigma$ is determined through
a non-linear least squares fits of the functional form
$\exp(-t/\tau)\cos(2 \pi t/T_{\sigma})$ to the oscillation data
($\tau$ is the characteristic damping time and $t$ is the hold
time). For $12 \le U_f \le 32 \, E_R$ we observe a monotonic
increase in $T_\sigma$ as $U_f$ is increased.  The observed period
shows a much weaker dependence on $U_f$ than, for example, the
characteristic period associated with the GPE predicted quadrupole
breathing mode \cite{stringari} (solid line in Fig. 3a).

On the other hand, TWA predicts a weak dependence of $T_{\sigma}$
on $U_f$, in agreement with observations.  TWA periods are
inferred from the time required for $D(t)$ to reach its maximum
value (see Fig.~\ref{figure3}b) but are scaled by an overall
factor of 2.1 in Fig.~\ref{figure3}a (dashed line). This scaling
factor may have its origin in the breakdown of TWA at longer
times.  In Fig.~\ref{figure3}c we represent the observed
frequencies in terms of effective phonon wave-vectors
$q_\mathrm{eff}$ through the (translationally invariant lattice)
phonon dispersion relation \cite{bogo},

\begin{equation}
\hbar\omega_q = \sqrt{4\gamma
\sin^2(\frac{qa}{2})[2N_ig\beta+4\gamma \sin^2(\frac{qa}{2})]}.
\end{equation}
Here $a = \lambda/2$ is the lattice period.  We find the value $q=
q_\mathrm{eff}$ such that $\omega_{q}$ is equal to the observed
frequency $2 \pi/T_{\sigma}$.  For sufficiently large $U_f$ the
observed frequencies correspond to short wavelength excitations,
whose frequency is proportional to the generalized Josephson
frequency.  For our parameters, this frequency is much faster than
the tunnelling frequency which has been observed to determine the
time scale for the onset of coherence in lattice systems with low
filling factor \cite{Bloch}. It is interesting to note that an
exponential fit to the data extrapolates to  $4 \pi/q_\mathrm{eff}
\lambda = 1$ site at 55 $E_R$, a lattice depth close to the MI
cross-over. Fig.~\ref{figure3}d shows the measured damping
coefficient $\tau$.  As qualitatively expected, faster damping
rates are observed with increased $U_f$ \cite{Tolya,Ehud}. Future
studies of the damping rate as $U_f$ is tuned near the
superfluid-MI phase transition may provide interesting insight
into the quantum critical regime \cite{Zurek, Ehud}.

\begin{figure}[htb!]
\includegraphics[scale = .4, bb = 0 20 650
800]{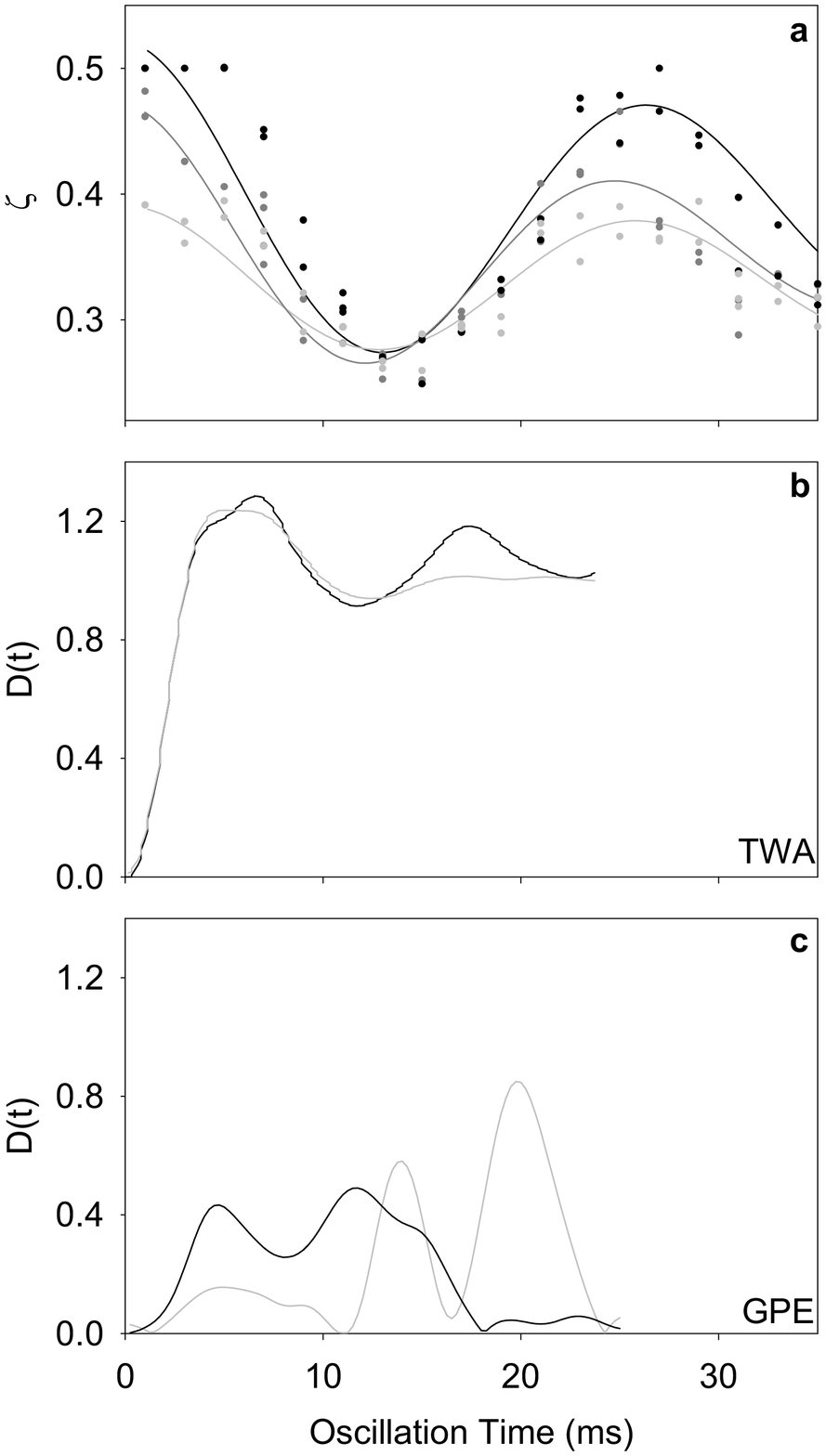} \caption{ \newline {\protect\small a)
Phase variance oscillations for $U_0$ = 80 (black), 63 $E_R$ (dark
grey) and $50$ $E_R$ (light grey) with $U_f$ = 17 $E_R$. b)
Numerical model of the lattice sequence using the TWA for $U_0$ =
80 $E_R$ (black) 50 $E_R$ (grey) c) Numerical model simulation of
$D(t)$ for lattice sequence in Fig.~\ref{figure2}a using GPE for
$U_0$ = 80 $E_R$ (black) and 50 $E_R$ (grey). }} \label{pofig4}
\end{figure}

Due to non-adiabaticity in the state preparation sequence, we also
investigate the possible effect of remnant phase coherence in the
initial state on the observed dynamics.  We do this by studying
the dependence of the oscillations on $U_0$, keeping $U_f$ fixed.
If the dynamics are driven largely by quantum fluctuations, we
expect dynamics to be independent of $U_0$ for large $U_0$ (where
the associated phase variance approaches 2$\pi$). We find
$\omega_{\sigma}/2\pi = 37.6 \pm 1.3$, $39.7 \pm 1.6$ and $38.91
\pm 1.6$ Hz, for $U_0 = 80$,  $63$ and $50$ $E_R$ respectively
with $U_f = 17$ $E_R$ (Fig. \ref{pofig4}a). These observations are
consistent with the TWA analysis, where $D(t)$ is shown in
Fig.~\ref{pofig4}b for $U_0 = 80$ $E_R$ and $50$ $E_R$.  For
comparison, we show numerical solutions to GPE, which we obtain by
integrating Eqs. 2 over the entire experimental sequence including
the state preparation, taking $\phi_j = 0$ as the initial
condition.  In contrast to our observations and the TWA, these
solutions show a strong dependence on the value of $U_0$ (Fig.
\ref{pofig4}c).  Finally, we independently assess the role of
non-adiabaticity through investigation of exact solutions to the
two well model for conditions of our experiments. In all cases, we
find the number fluctuations of the initial state to be at the 1
atom per lattice site level and observe no dependence of the
observed frequency on residual number fluctuations.

In summary, we have shown that the TWA can accurately model the
non-equilibrium dynamics for the soft boson lattice system.  In
future work we will seek to study these dynamics in the quantum
critical regime as well as to exploit them to demonstrate
sub-shot-noise interferometry.


\begin{thebibliography}{99}
\bibitem{integrals} $\gamma \equiv \int d^3r \phi_i(r)\left(-\frac{\hbar^2}{%
2m}\nabla^2 + U(r)\right) \phi_j(r)$ and $\beta_i \equiv \int d^3r
\phi_i^2 \phi_i^2$, where $\phi_{i,j}$ are nearest neighbor
single-particle states, and $U(r)$ is the lattice potential. In
the Thomas-Fermi regime in the radial direction, we find $g\beta_i
= [m \omega_{z0} \omega_{\perp}^4 \hbar^3 a^2/2 \pi {N_i}^2]^{1/4}$ with $%
\omega_{\perp}$ and $\omega_{z0}$ the radial and longitudinal
oscillation frequencies characteristic of an individual lattice
site.
\bibitem{lattice} Lattice depth is calibrated by resonantly heating atoms
from the confining lattice potential. The peak lattice depth is
inferred from measurements of atom loss as a function of
modulation frequency and has a calibration error of $\sim 20{\%}$.
\end{thebibliography}
\end{document}